\documentclass[twocolumn,english,prl]{revtex4}
\usepackage[T1]{fontenc}
\usepackage[latin9]{inputenc}
\usepackage{bm}
\usepackage{amsmath}
\usepackage{amssymb}
\usepackage{graphicx}
\usepackage{esint}

\makeatletter

\@ifundefined{textcolor}{}
{%
 \definecolor{BLACK}{gray}{0}
 \definecolor{WHITE}{gray}{1}
 \definecolor{RED}{rgb}{1,0,0}
 \definecolor{GREEN}{rgb}{0,1,0}
 \definecolor{BLUE}{rgb}{0,0,1}
 \definecolor{CYAN}{cmyk}{1,0,0,0}
 \definecolor{MAGENTA}{cmyk}{0,1,0,0}
 \definecolor{YELLOW}{cmyk}{0,0,1,0}
}

\usepackage{float}\usepackage{bm}

\usepackage{babel}

\makeatother

\usepackage{babel}
\begin{document}

\preprint{This line only printed with preprint option}

\title{Critical magnetic field of ultrathin superconducting films and interfaces}

\author{Gertrud Zwicknagl}

\email{g.zwicknagl@tu-bs.de}

\selectlanguage{english}%

\affiliation{Institut für Mathematische Physik, Technische Universität Braunschweig,
38106 Braunschweig, Germany}

\author{Peter Fulde}

\email{fulde@pks.mpg.de}

\selectlanguage{english}%

\affiliation{Max-Planck-Institut für Physik komplexer Systeme, 01187 Dresden,
Germany}

\author{Simon Jahns}

\email{s.jahns@tu-braunschweig.de}

\selectlanguage{english}%

\affiliation{Institut für Mathematische Physik, Technische Universität Braunschweig,
38106 Braunschweig, Germany}
\begin{abstract}
We derive an analytic expression for the temperature dependent critical
magnetic field parallel to ultrathin superconducting films with Rashba
spin-orbit interaction. Thereby we cover the range from small to large
spin-orbit interactions $\lambda$ compared with the gap parameter
$\Delta_{0}$. We find that at a critical spin-orbit energy $\lambda_{c}$
a first-order phase transition takes place at which the pairing momentum
of the Cooper pairs changes discontinuously. We speculate that this
might give raise to new phenomena. With increasing $\lambda/\Delta_{0}$
the pair formation changes from interband to intraband pairing. For
$\lambda>\lambda_{c}$ a dimensional cross-over of the critical field
from two to one dimension is taking place. 
\end{abstract}

\date{\today}

\maketitle
In recent years the manufactoring of controlled ultrathin superconducting
films has made impressive progress. Important examples are monoatomic
or monomolecular layers on a substrate, \cite{Zhang10,Sekihara13},
superconducting layers in a superlattice \cite{Mizukami11,Shimozawa16},
or superconducting interfaces and surfaces \cite{Gariglio15,Ge15}
. They have in common the absence of inversion symmetry and hence
the presence of Rashba-type spin-orbit energy $\lambda$ \cite{BauerSigristBook}.
The latter can be modified to some extent by varying the thickness,
the number of layers, or by applying an electric voltage.

Spin-orbit interactions and their effects on superconductivity were
considered shortly after the development of the BCS theory \cite{Ferrell59,Anderson59,DeGennes64}.
However, in distinction to the Rashba-type of interaction the majority
of studies focused on impurity or surface scattering which does not
conserve momentum. As a consequence of its invariance under time reversal,
spin-orbit scattering off impurities does not affect the superonducting
transition temperatures of isotropic s-wave superconductors. However,
it leads to a finite spin-susceptibiliy of the ground state. When
the spin populations of electrons become unbalanced by an applied
magnetic field, Cooper pairing is quite different when momentum is
conserved or violated by spin-orbital interaction processes. Therefore
the critical magnetic field caused by the Zeeman effect varies strongly
in the two cases. Spin-orbit scattering off impurities derives from
the atomic potentials of heavier elements. When the latter, however,
are sitting on regular lattice sites one has to account for a periodic
spin-orbit interaction which can be rather strong compared to the
characteristic energies of a superconductor and which is consequently
accounted for in terms of the normal-state quasiparticles and their
interactions. The consequences of periodic spin-orbit interaction
were discussed in a seminal paper by Bulaevskii et al. \cite{Bulaevskii76}. 
\begin{figure}
\begin{centering}
\includegraphics[width=0.8\columnwidth]{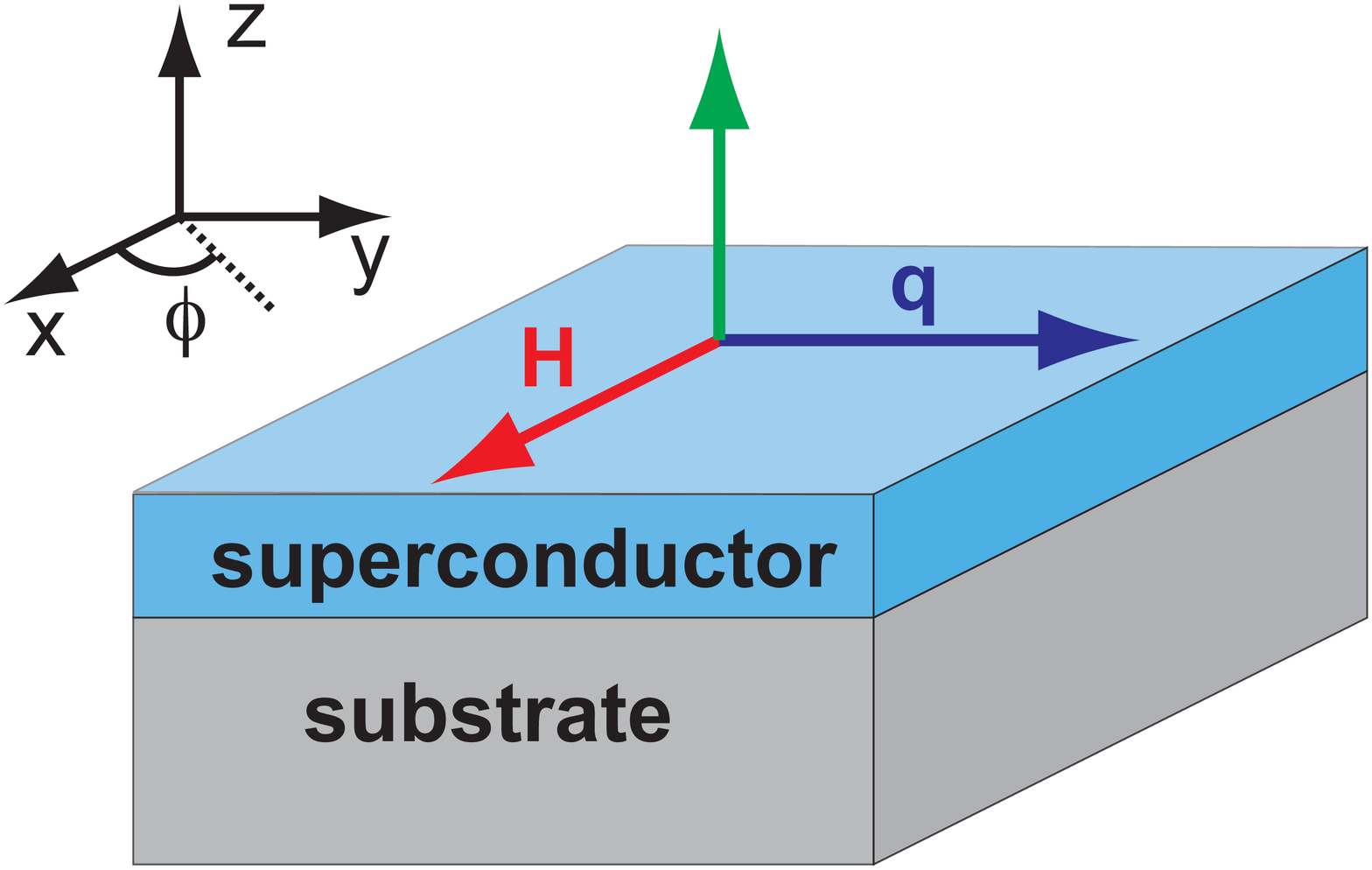}
\par\end{centering}

\vspace{0.5cm}

\centering{}\includegraphics[width=0.8\columnwidth]{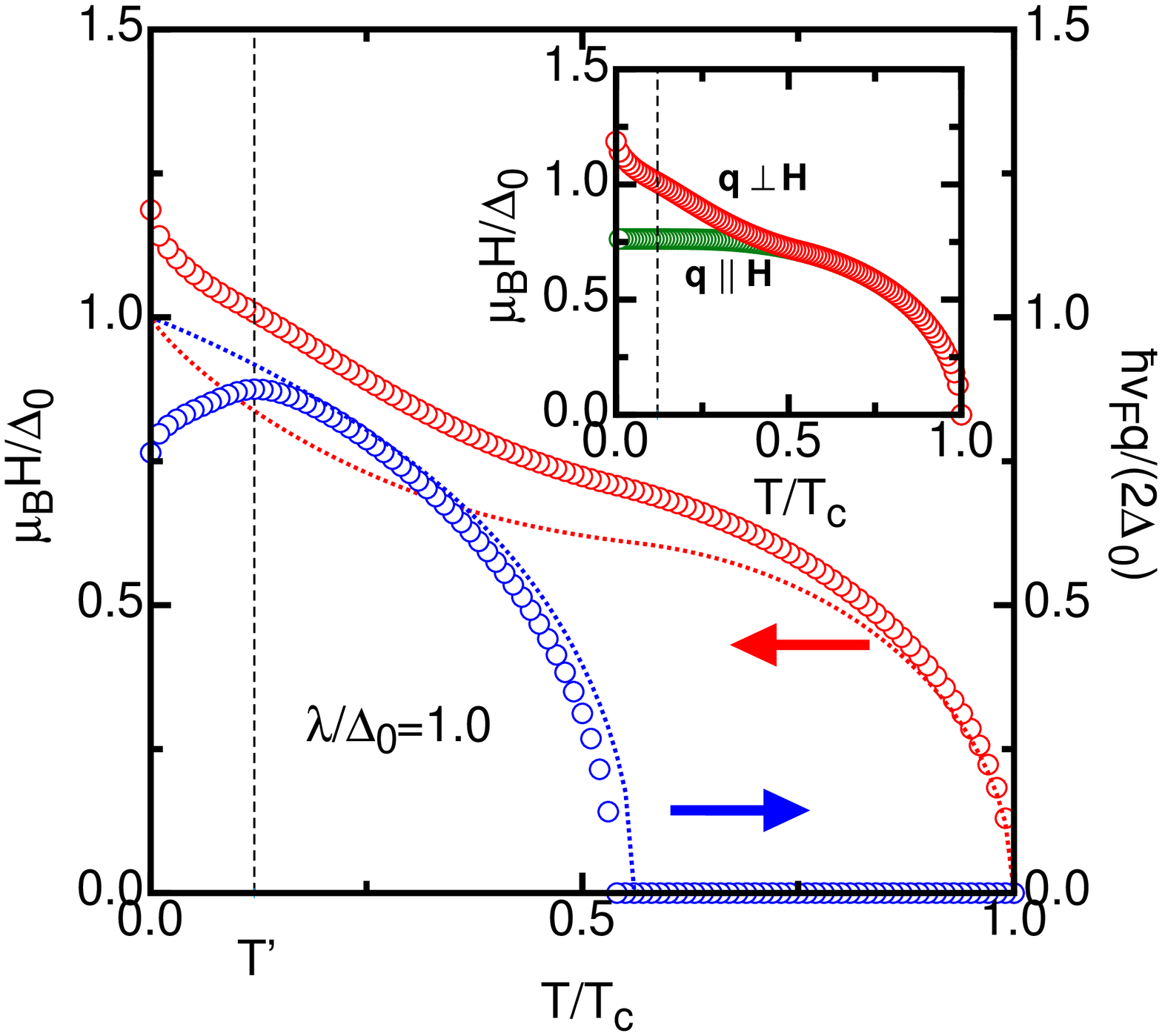}
\caption{Influence of Rashba spin-orbit interaction $\lambda=\Delta_{0}$ on
the normal-to-superconductor transition of an ultra-thin film in an
in-plane magnetic field. The geometry is explained in the upper panel.
Lower panel: The upper critical field $H_{c2}\left(T\right)$ (red
dots) is enhanced over the universal BCS-prediction for $\lambda=0$
(dotted red line). The momentum $q\left(T\right)$ (blue dots) of
the Cooper pairs forming at $H_{c2}\left(T\right)$ is (slightly)
reduced compared to the case $\lambda=0$ (dotted blue line). The
tri-critical point $T^{*}$ at which the normal phase merges with
the superconducting phases with $q=0$ and $q\neq0$ is shifted to
lower temperatures. The variation $q\left(T\right)$ is non-monotonic
with a maximum at $T'<T^{*}$. The hypothetical critical magnetic
field for Cooper pairs with ${\bf q}\parallel{\bf H}$ is shown in
the inset for comparison (green dots). \label{fig:AnomaliesVariationWithTLambda}}
\end{figure}

Here we focus on inhomogeneous superconducting phases caused by an
in-plane magnetic field in a quasi-2D superconductor with Rashba interaction
(see Figure \ref{fig:AnomaliesVariationWithTLambda}). They are due
to the Zeeman effect and an analogue of those caused by a magnetic
field acting on the electron orbits rather than on the spins (Abrikosov
lattices). 

We assume that the attractive interaction leading to the formation
of Cooper pairs is the same as in the corresponding bulk superconductor
which we use as reference system. To characterize the superconducting
properties of ultra-thin layers, we measure the energies and lengths
in units of the low-temperature energy gap $\Delta_{0}$ and the coherence
length $\xi_{0}=\frac{\hbar v_{F}}{\pi\Delta_{0}}$ of the reference
system. Here $\hbar$ is Planck's constant and $v_{F}$ denotes the
Fermi velocity in the normal state. For simplicity, we consider a
single-component spin-singlet superconductor for which the order parameter
$\Delta\left(\text{{\bf R}}\right)\mathcal{Y}\left(\hat{{\bf k}}\right)$
can be factorized into a spatially dependent complex amplitude and
a momentum-space basis function where $\hat{{\bf k}}$ is the direction
of the Fermi momentum. The basis function is normalized $\left\langle \left|\mathcal{Y}\left(\hat{{\bf k}}\right)\right|^{2}\right\rangle _{FS}=1$
where $\left\langle \ldots\right\rangle _{FS}$ denotes the angular
Fermi surface average.

The population imbalance generated by an in-plane magnetic field of
magnitude $H$ will depend on the relative strengths of the spin-orbit
interaction $\lambda$ for the quasiparticles at the Fermi energy
and the Zeeman energy $h=\frac{1}{2}g\mu_{B}H$ where $g=2$ and $\mu_{B}$
denote the gyromagnetic ratio and the Bohr magneton, respectively. 

There exists already a considerable body of work on superconductivity
of unbalanced populations in the absence of inversion symmetry \cite{Kaur05,Gorkov01,Agterberg07,Loder15,Zheng14,Zhou14,ZhangW13}.
It was partially motivated by expectations that experiments on ultracold
fermionic atoms on optical lattices could test theoretical predictions
\cite{Zheng14,Zhou14,ZhangW13}. In particular, phase diagrams were
calculated numerically for various limiting cases. 

The present paper describes the results of a microscopic theory of
superconducting films with population imbalance which are subject
to Rashba spin-orbit interaction. The full range of spin-orbit interaction
energies $\lambda$ is considered, i.e., from $\lambda\ll\Delta_{0}$
to $\lambda\gg\Delta_{0}$ where $\Delta_{0}$ is the superconducting
gap. Of particular interest are the variation with temperature of
the upper critical field $H_{c2}$ below which the normal phase becomes
unstable. 

\begin{flushleft}
Figure \ref{fig:AnomaliesVariationWithTLambda} illustrates central
results of the present paper for an isotropic s-wave superconductor.
The subtle interplay of imbalance created by the magnetic field and
the Rashba spin-orbit interaction gives rise to an enhancement of
the upper critical field curve $H_{c2}\left(T\right)$ over the universal
BCS curve. This enhancement is dramatic for $\lambda\gg\Delta_{0}$
\cite{SupplHc2Enhancement}. In addition, the tri-critical point $T^{*}$
at which the normal phase merges with the superconducting phases with
pairing momentum $q=0$ and $q\neq0$ is shifted to lower temperatures
\cite{SupplTriCrit}. Due to the coupling of spin and orbital motion,
the direction of the modulation vector is fixed perpendicular to the
magnetic field direction ${\bf q}\left(T\right)\perp{\bf H}$. 
\par\end{flushleft}

A more striking feature of the upper critical field curve is the existence
of a temperature $T'<T^{*}$ below which the phase boundary steepens
as function of temperature. At this temperature $T'$, the magnitude
of the modulation vector is maximal decreasing for decreasing temperatures
$T<T'$. This non-monotonic variation of $q\left(T\right)$ for $\lambda$
comparable to $\Delta_{0}$ is a feature of the theory presented here.

The theory leading to the results in Figure \ref{fig:AnomaliesVariationWithTLambda}
is based on the quasiclassical method and will be discussed in an
extended version of this paper. The idea behind it is more simple.
We must generalize the magnetic-field induced imbalance to systems
with spin-orbit interaction. 

In the absence of SO interaction the upper critical field is given
in terms of digamma functions $\psi\left(z\right)$ by \cite{SaintJamesBook}

\begin{align}
\ln\frac{T}{T_{c}} & =\text{\ensuremath{\psi\left(\frac{1}{2}\right)}-\ensuremath{\left\langle \mathrm{Re}\,\psi\left(\frac{1}{2}+i\frac{h+\frac{\hbar}{2}{\bf v}_{F}\cdot{\bf q}}{2\pi k_{B}T}\right)\left|\mathcal{Y}\left(\hat{{\bf k}}\right)\right|^{2}\right\rangle }}_{FS}\nonumber \\
0 & =\frac{\partial}{\partial q}\left\langle \mathrm{Re}\,\psi\left(\frac{1}{2}+i\frac{h+\frac{\hbar}{2}{\bf v}_{F}\cdot{\bf q}}{2\pi k_{B}T}\right)\left|\mathcal{Y}\left(\hat{{\bf k}}\right)\right|^{2}\right\rangle _{FS}\quad.\nonumber \\
\label{eq:Hc2Zeeman}
\end{align}
These conditions evaluated for an isotropic s-wave superconductor
produce the dotted lines in Figure \ref{fig:AnomaliesVariationWithTLambda}.
Here $h+\frac{\hbar}{2}{\bf v}_{F}\cdot{\bf q}=\frac{1}{2}\left|\text{\ensuremath{\epsilon}}_{\uparrow}\left({\bf {\bf k}}+\frac{{\bf q}}{2}\right)-\text{\ensuremath{\epsilon}}_{\downarrow}\left(-{\bf {\bf k}}+\frac{{\bf q}}{2}\right)\right|$
is directly related to the energy difference between states forming
a Cooper pair with finite pairing momentum ${\bf q}$.

Due to spin-orbit interaction, spin ceases to be a good quantum number
and, as a result, the pair density will generally contain singlet
and triplet contributions. The short-ranged attraction, however, which
we assume to be the same as in the bulk reference superconductor leads
to a spin-singlet order parameter which is a coherent superposition
of inter- and intraband pairs To discuss the consequences for the
upper critical field we start with the linear equation for the existence
of a superconductivity gap function
\begin{equation}
\Delta\left({\bf R}\right)=\int d^{3}R'\, K\left({\bf R},{\bf R'}\right)\,\Delta\left({\bf R}'\right)\label{eq:LinSelfconGeneral}
\end{equation}
where the non-local kernel $K\left({\bf R},{\bf R}'\right)$ has to
be calculated from a microscopic theory. It depends on the properties
of the quasiparticles in the normal state which are described by the
Hamiltonian \cite{Kaur05} 
\begin{eqnarray}
H_{0} & = & \sum_{{\bf k}ss'}\left(\hat{H}_{0}\left({\bf k}\right)\right)_{ss'}c_{{\bf k}s}^{\dagger}c_{{\bf k}s'}\nonumber \\
\hat{H}_{0}\left({\bf k}\right) & = & \xi_{{\bf k}}\hat{{\bf 1}}-\frac{g}{2}\mu_{B}{\bf H\cdot\hat{\bm{\sigma}}}+\lambda\left({\bf e}_{z}\times\hat{{\bf k}}\right)\cdot\hat{\bm{\sigma}}\label{eq:H0}
\end{eqnarray}
with the $2\times2$-unit matrix $\hat{{\bf 1}}$, the Pauli matrix
vector $\hat{\bm{\sigma}}=\left(\hat{\sigma}_{1},\hat{\sigma}_{2},\hat{\sigma}_{3}\right)$,
and the kinetic energy of the quasiparticles $\xi_{{\bf k}}=v_{F}\left(k-k_{F}\right)$.
Throughout the calculations we choose the z-direction as quantization
axis. The constant $\lambda$ is the spin-orbit energy of the states
at the Fermi energy.

The phase boundary between the normal state and the single-${\bf q}$-phase
$\Delta\left({\bf R}\right)\sim e^{i{\bf q\cdot{\bf R}}}$ with $\left|{\bf q}\right|\ll k_{F}$
is determined by\cite{deGennesBook,SaintJamesBook} 
\begin{equation}
K\left({\bf q};H,T\right)-K\left(0;0,T_{c}\right)=0\quad;\quad\nabla_{{\bf q}}K\left({\bf q};H,T\right)=0\label{eq:Hc2General}
\end{equation}
 where we eliminated the BCS-coupling constant and the cut-off in
favor of the transition temperature $T_{c}$. Replacing $\sum_{{\bf k}}\ldots\to N\left(0\right)\left\langle \int d\xi_{{\bf k}}\ldots\right\rangle _{FS}$
yields for the first condition

\begin{widetext}

\begin{equation}
\ln\frac{T}{T_{c}}=k_{B}T\sum_{\epsilon_{n}}\left(\left\langle \int d\xi_{{\bf k}}\frac{1}{2}\mathrm{Tr}\left\{ \hat{\sigma}_{2}\hat{G}_{0}\left({\bf k}+\frac{{\bf q}}{2};i\epsilon_{n}\right)\hat{\sigma}_{2}\hat{G}_{0}^{*}\left({\bf -k}+\frac{{\bf q}}{2};i\epsilon_{n}\right)\right\} \left|\mathcal{Y}\left(\hat{{\bf k}}\right)\right|^{2}\right\rangle _{FS}-\frac{\pi}{\left|\epsilon_{n}\right|}\right)\label{eq:KernelWithGreenFunctions}
\end{equation}
with the normal-state Green's functions $\hat{G}_{0}\left({\bf k}+\frac{{\bf q}}{2};i\epsilon_{n}\right)=\left[\left(i\epsilon_{n}-\frac{\hbar}{2}{\bf v}_{F}\left(\hat{{\bf k}}\right)\cdot{\bf q}\right)\hat{1}-\hat{H}_{0}\left({\bf k}\right)\right]^{-1}$
at the Matsubara frequencies $\mbox{{\ensuremath{\epsilon_{n}}=\ensuremath{\pi k_{B}}T\ensuremath{\left(2n+1\right)}}}$.
Evaluating the expressions in Eq. (\ref{eq:KernelWithGreenFunctions})
yields
\begin{equation}
\ln\frac{T}{T_{c}}=\psi\left(\frac{1}{2}\right)-\left\langle \left[\left|u_{{\rm inter}}\right|^{2}\textrm{Re }\psi\left(\frac{1}{2}+i\frac{W_{inter}+\frac{\hbar}{2}{\bf v}_{F}\cdot{\bf q}}{2\pi k_{B}T}\right)+\left|u_{{\rm intra}}\right|^{2}\textrm{Re }\psi\left(\frac{1}{2}+i\frac{W_{intra}+\frac{\hbar}{2}{\bf v}_{F}\cdot{\bf q}}{2\pi k_{B}T}\right)\right]\left|\mathcal{Y}\right|^{2}\right\rangle _{FS}\label{eq:Hc2WithSOFiniteT}
\end{equation}
\end{widetext}with
\begin{eqnarray}
W_{inter/intra}^{2}\left(\hat{{\bf k}}\right) & = & \frac{1}{2}\left(\lambda^{2}+h^{2}\right)\nonumber \\
 &  & \pm\frac{1}{2}\sqrt{\left(\lambda^{2}+h^{2}\right)^{2}-4h^{2}\lambda^{2}\hat{k}_{y}^{2}}\label{eq:EnDiffIntraVsInterBand}
\end{eqnarray}
\begin{align}
\left|u_{inter/intra}\left(\hat{{\bf k}}\right)\right|^{2} & =\frac{1}{2}\left\{ 1\pm\frac{h^{2}-\lambda^{2}}{\sqrt{(\lambda^{2}+h^{2})^{2}-4h^{2}\lambda^{2}\hat{k}_{y}^{2}}}\right\} \quad.\label{eq:WeightsIntraVsInterBandPairs}
\end{align}

This is the central result of the present paper. It is a straight-forward
generalization of Eq. (\ref{eq:Hc2Zeeman}). The upper critical magnetic
field is determined by half the energy differences $W_{inter/intra}+\frac{\hbar}{2}{\bf v}_{F}\cdot{\bf q}$
of the quasiparticle states ${\bf k}+\frac{{\bf q}}{2}$ and $-{\bf k}+\frac{{\bf q}}{2}$
taken from the same (intra) and different (inter) bands, respectively.
The contributions of the two types of pairing are weighted by $\left|u_{inter/intra}\right|^{2}$.

\begin{figure}
\includegraphics[width=0.5\columnwidth]{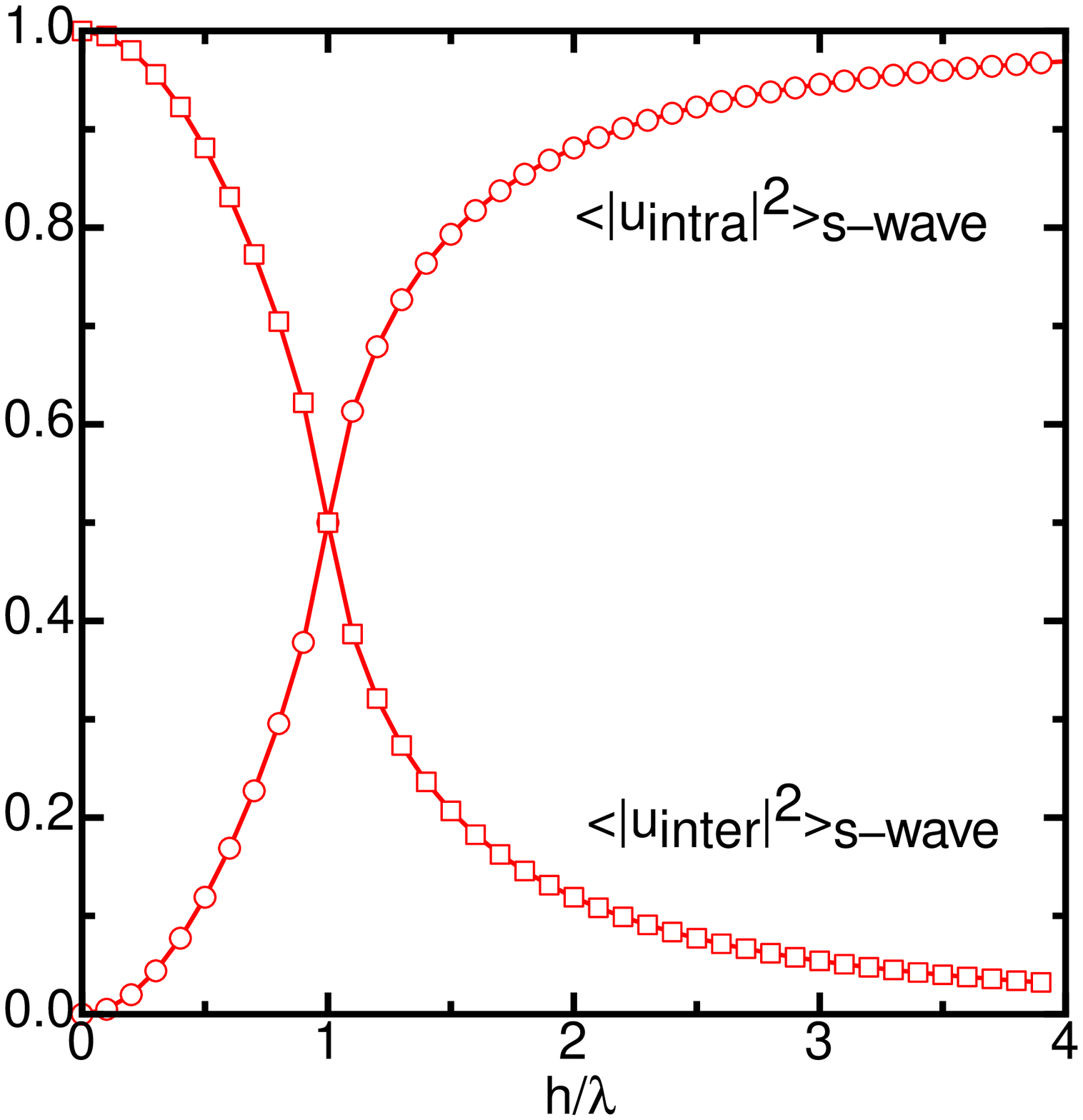}\includegraphics[width=0.5\columnwidth]{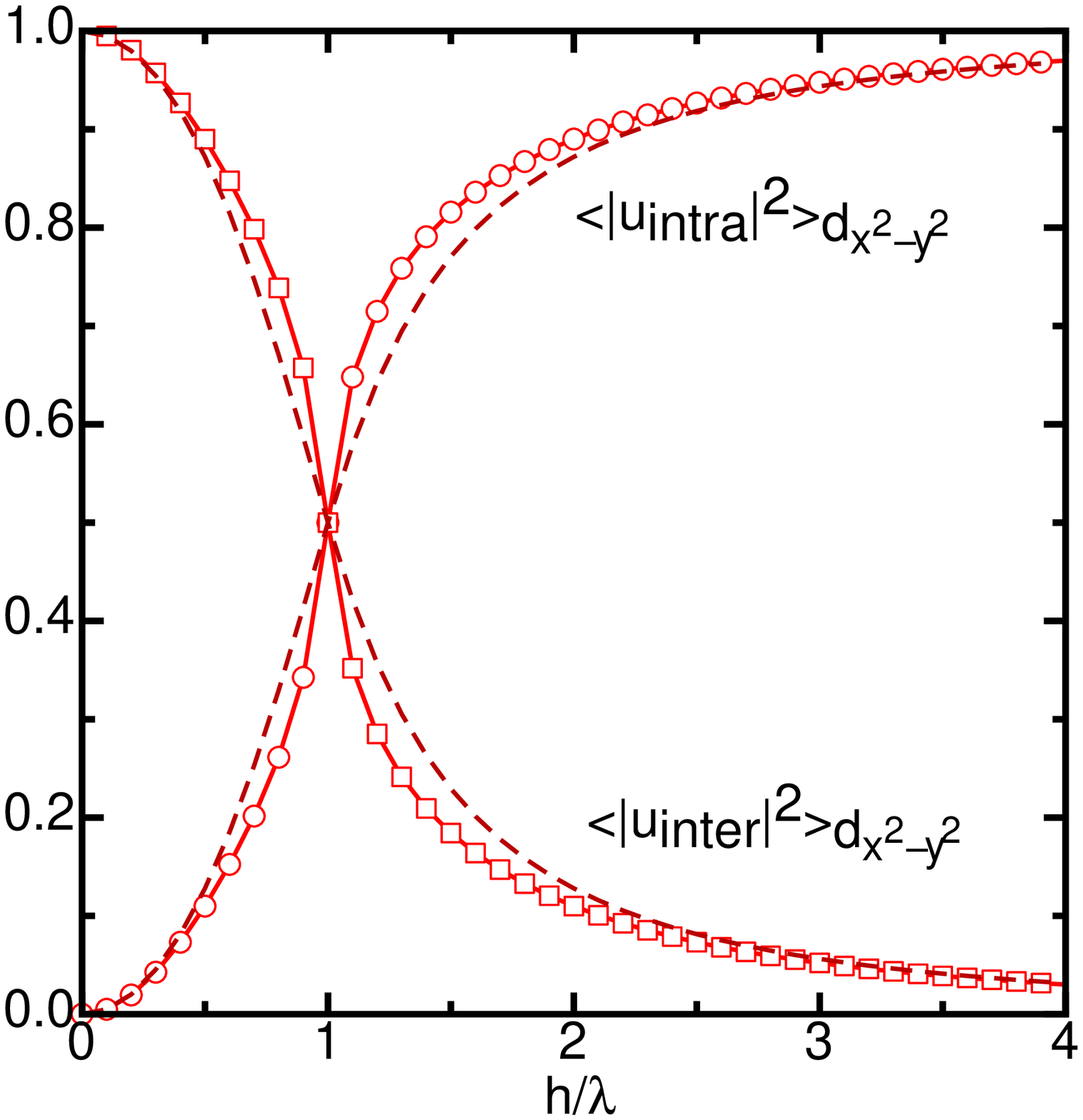}
\caption{Averaged weights of interband (squares) and intraband (dots) contributions
from Eq. (\ref{eq:WeightsIntraVsInterBandPairs}) to isotropic s-wave
(left panel) and $d_{x^{2}-y^{2}}$-pair states with the in-plane
magnetic field in anti-nodal directions (right panel). The case of
magnetic field in nodal direction (dashed lines) is included for camparison.
For low magnetic fields, the intraband contributions dominate while
the interband pairs dominate for large magnetic fields. For its numerical
evaluation see \cite{SupplAveragedWeights}.}

\label{fig:PairDecomposition} 
\end{figure}

The brackets $\langle\dots\rangle_{FS}$ denote an angular average
over the Fermi surfaces of the two spin-orbit split bands. Here to
leading order in $\lambda/E_{F}$, the difference $k_{F\pm}=k_{F0}(1\pm\lambda/E_{F})$
can be neglected when the angular averages are taken. The averaged
weights for the two types of pairing are plotted in Fig. \ref{fig:PairDecomposition}
for s-wave as well as d-wave pairing when the field is in nodal and
anti-nodal direction, respectively. The averaging integrals can be
done analytically and lead to elliptic integrals. 

Because of the neglect of the difference $k_{F\pm}$ in averaging,
the solutions of Eq. (\ref{eq:Hc2WithSOFiniteT}) are still degenerate
with respect to $\pm{\bf q}$. In reality this degeneracy is slightly
split depending on the size of $\lambda$. Yet, this has little effect
on $h_{c2}\left(T\right)$. Lifting the $\pm{\bf q}$ degeneracy implies
a ground state with a finite spin current yet vanishing charge current.
The depaired electrons have unbalanced spin populations in this case,
a topic dealt with in a separate investigation.

\begin{figure}
\includegraphics[width=0.7\columnwidth]{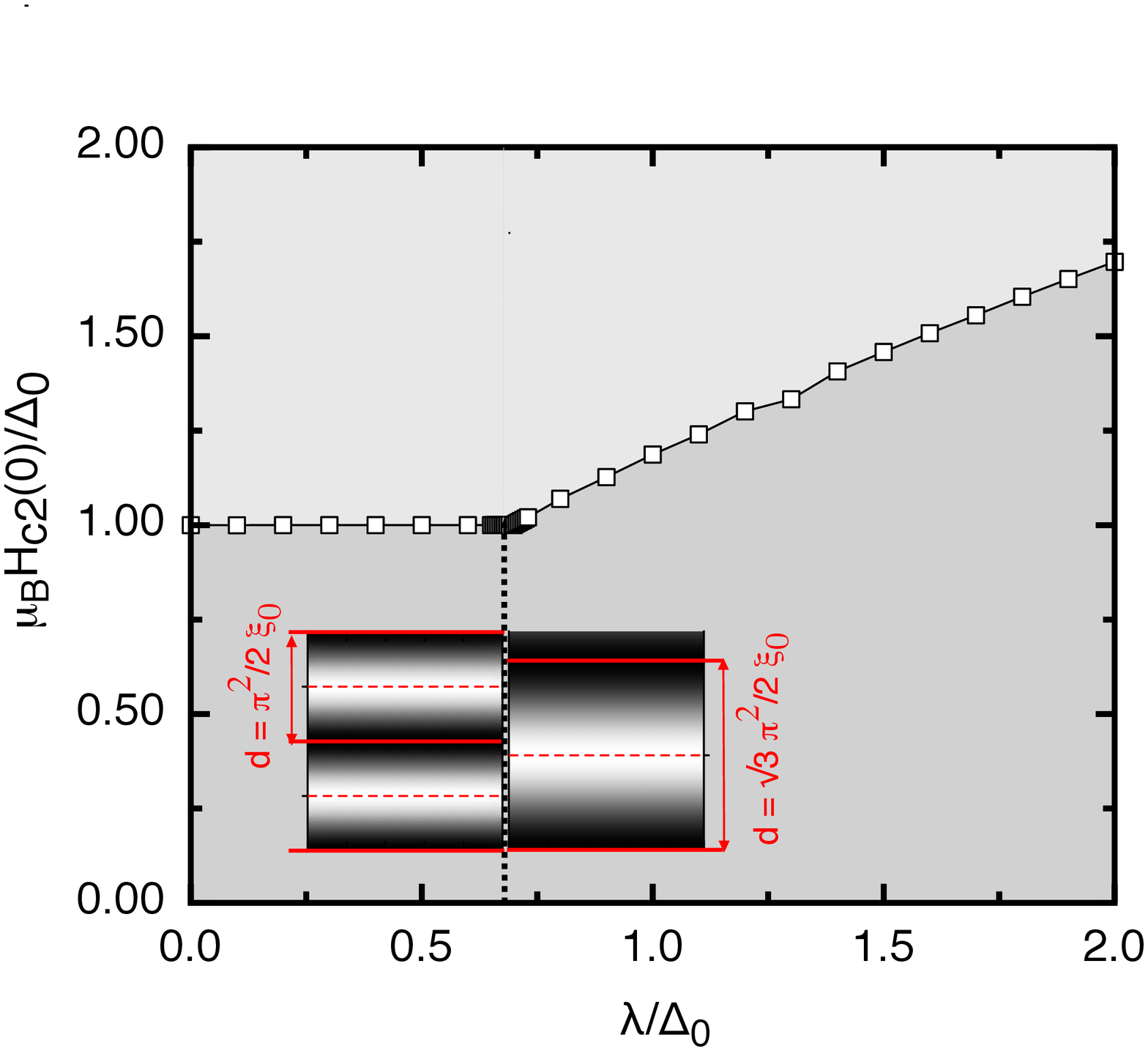}

\includegraphics[width=0.7\columnwidth]{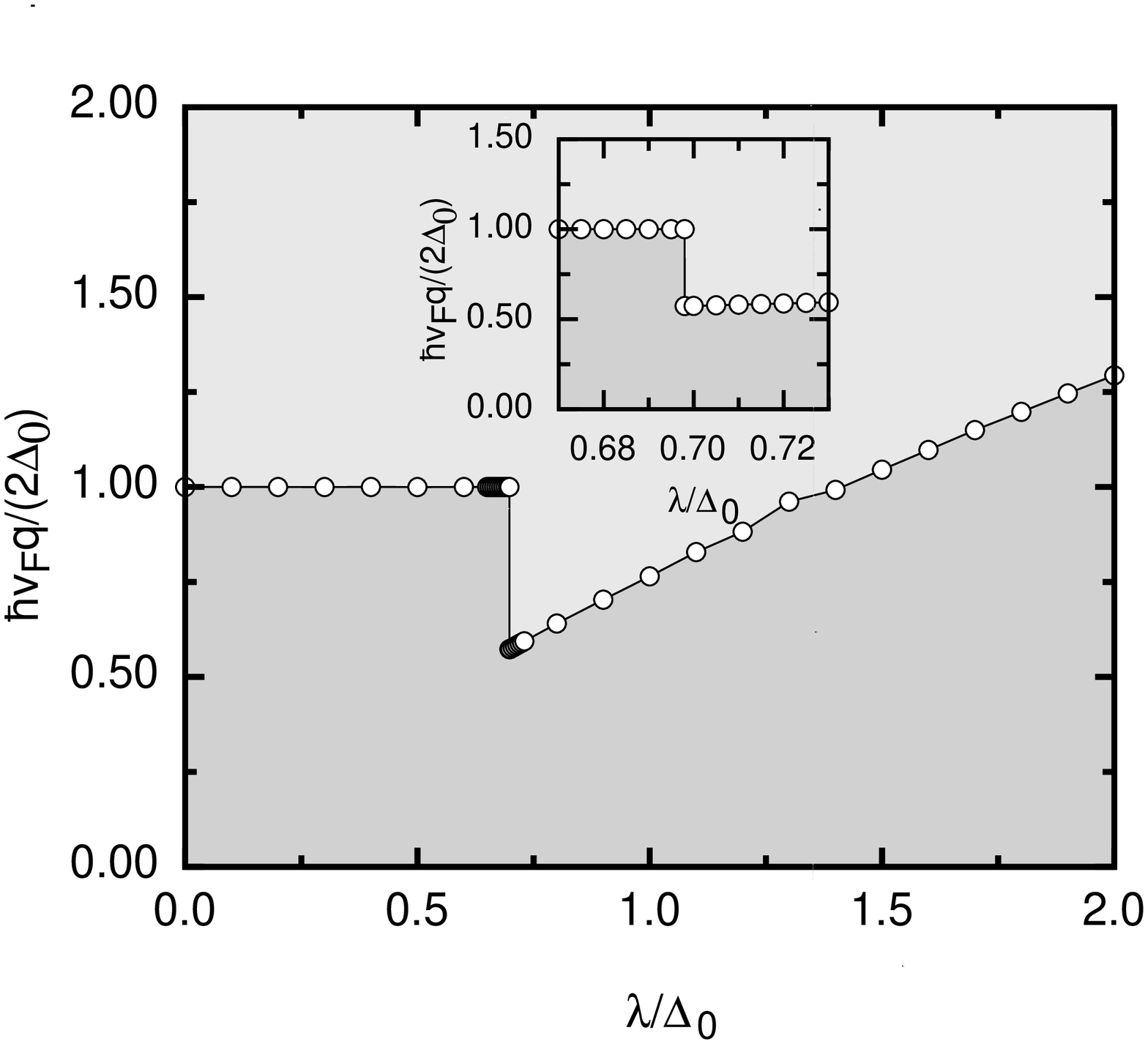}

\caption{Critical field $h/\Delta_{0}$ (upper panel) and modulation $Q=\hbar v_{F}q/(2\Delta_{0})$
with ${\bf q}$ perpendicular to the magnetic field as function of
$\lambda/\Delta_{0}$. At $\lambda_{c}/\Delta_{0}=1/\sqrt{2}$, a
discontinuous jump in pairing momentum from $Q=\Delta_{0}$ to $Q=\Delta_{0}/\sqrt{3}$
is taking place. }

\label{fig:ZeroTResults}
\end{figure}

Information on the ground state in the presence of a magnetic field
is obtained by reducing Eq. (\ref{eq:Hc2WithSOFiniteT}) to the zero-temperature
limit. The results are summarized in Figure \ref{fig:ZeroTResults}.

\begin{figure}
\includegraphics[width=0.9\columnwidth]{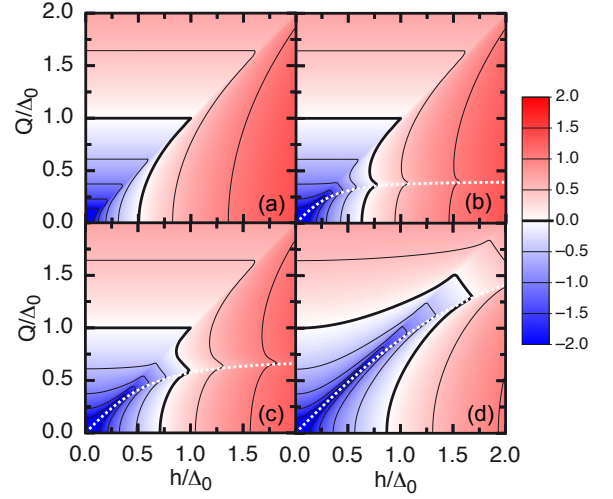}

\caption{Contour plot of the LHS of Eq. (\ref{eq:Hc2WithSOFiniteT}) for $Q=\hbar v_{F}q/(2\text{\ensuremath{\Delta}}_{0})$
vs. $h/\Delta_{0}$ at $T=0$. The $\lambda/\Delta_{0}$ values are
(a - d): 0, 0.4, $\lambda_{c}/\Delta_{0}=1/\sqrt{2}=$0.707 and 1.5.
The thick black lines show the solutions of the linearized self-consistency
equations. The dotted white lines correspond to the ``nesting''
condition for intra-band pairing $\hbar v_{F}q/2=Q=h\lambda/\sqrt{h^{2}+\lambda^{2}}$.
The second peaked solution gives a larger value of $h$ when $\lambda>\lambda_{c}=\Delta_{0}/\sqrt{2}$.}

\label{fig:EvolutionOfSecondSolution}
\end{figure}

We find that for $\lambda<\Delta_{0}/\sqrt{2}$, the value of $h_{c2}=\Delta_{0}$
remains essentially uneffected by $\lambda$, with $Q=\frac{\hbar v_{F}q}{2}=\Delta_{0}$
like for $\lambda=0$ \cite{Zwicknagl11b,Zwicknagl10}. However, at
a critical value of $\lambda_{c}=\Delta_{0}/\sqrt{2}$, a second superconducting
state with $Q=\Delta_{0}/\sqrt{3}$ yields the same critical field
$h=\Delta_{0}$ as does $Q=\Delta_{0}$ . It determines the superconducting
to normal transition for $\lambda>\lambda_{c}$ as it yields a higher
values of $h$. This is seen in Fig. \ref{fig:EvolutionOfSecondSolution}
where it can be noted that the state with $h=\Delta_{0}$ and $Q=\Delta_{0}$
continues to exist up to $\lambda\leq\Delta_{0}$ although it is unstable
for $\lambda>\lambda_{c}$. With increasing $\lambda>\Delta_{0}$
the self-consistent solutions go over into the ones found for a one-dimensional
system i.e., the g-factor becomes strongly anisotropic. The phase
transition at $\lambda_{c}$ found here and leading to a dimensional
cross-over is a new feature and has not previously been observed.
It is tempting to speculate, that the existence of two distinct modulation
vectors near $\lambda_{c}$ might give rise to novel phenomena. For
example, the tunneling density of states should change dramatically
near this point.

\begin{figure}
\includegraphics[width=0.4\columnwidth]{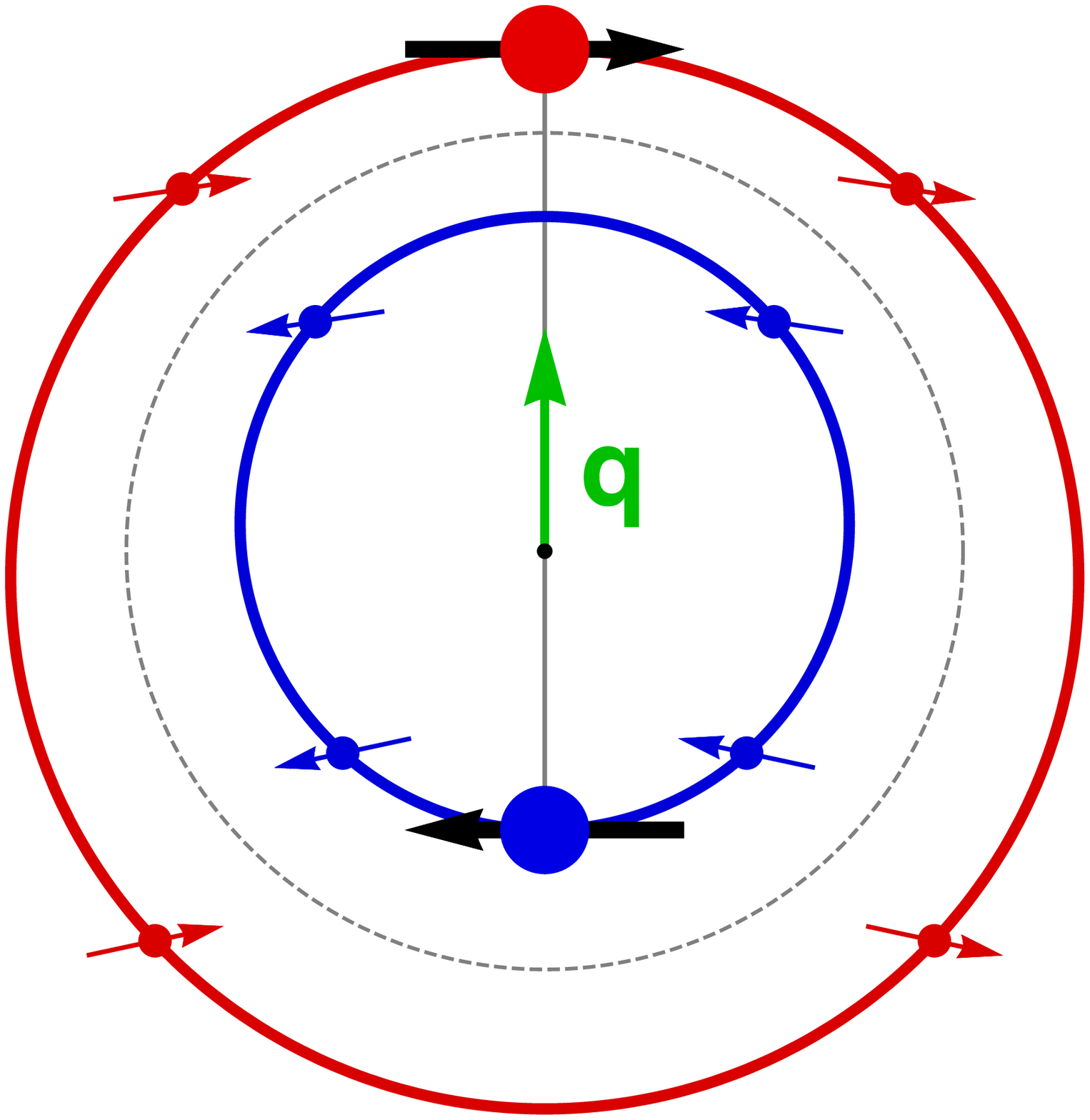}\hspace{1cm}\includegraphics[width=0.45\columnwidth]{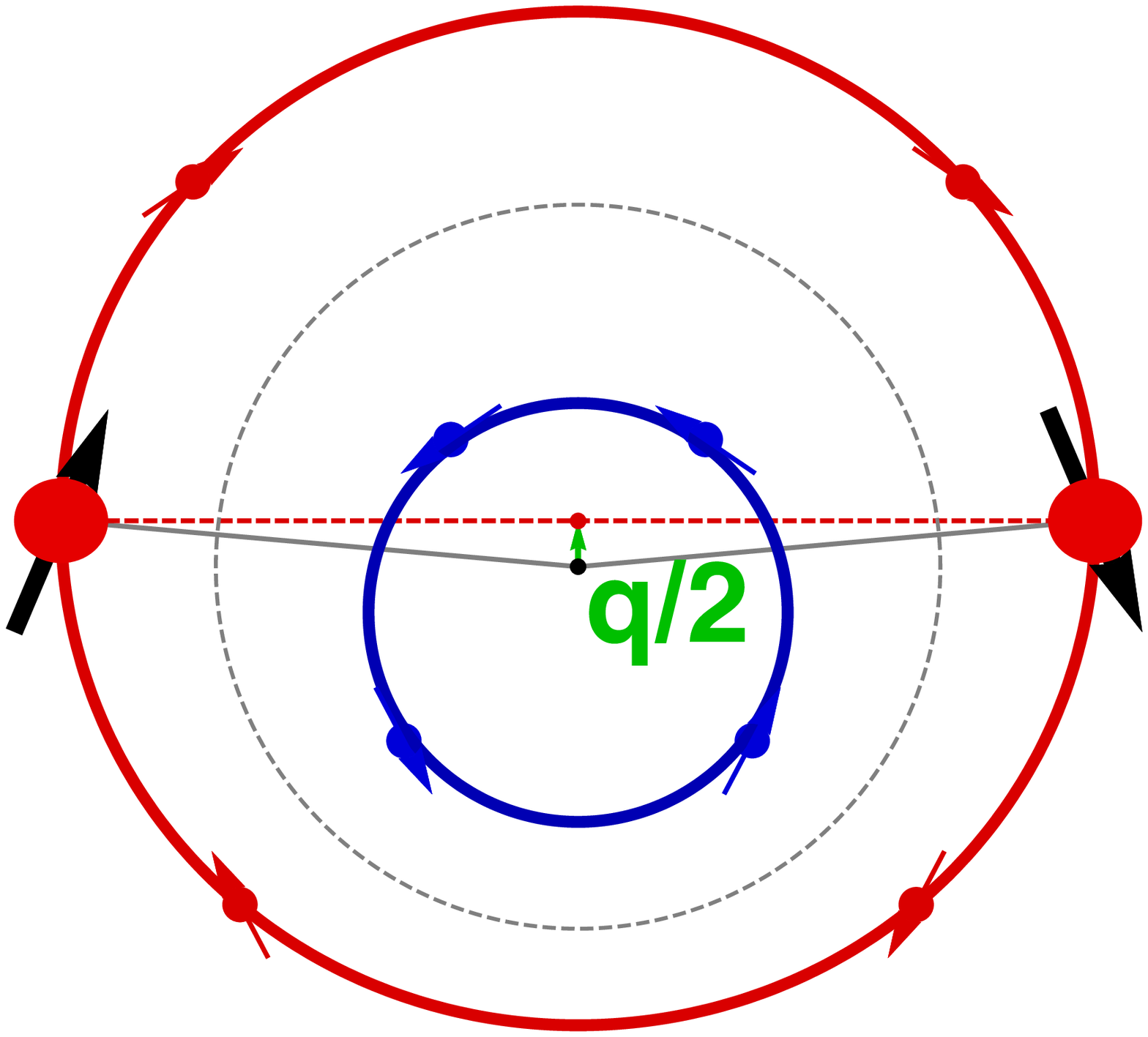}

\caption{Finite-momentum pairing in limiting cases. For $\lambda\ll h$ (left
panel) the dominant contribution comes from interband pairing and
we find the usual FFLO scenario. For $\lambda\gg h$ (right panel),
intraband pairing dominates and the pairing momentum is related to
the shift of the Fermi surfaces as explained in the text.}

\label{fig:TwoVectors}
\end{figure}

The existence of the two modulation vectors can be understood from
rather elementary consideration. The center-of-gravity momentum of
the Cooper pairs, ${\bf q}$, is selected so as to minimize depairing
due to imbalance. In the present case, the Cooper pairs contain contributions
from both inter- and intra-band pairs. For weak SO interaction, the
interband pairs dominate and we encounter the well-known FFLO scenario
of a quasi-2D superconductor with $h_{c2}(0)=\Delta_{0}$ and $Q=\Delta_{0}$.
As the strength of the SO interaction increases relative to the Zeeman
energy, the intraband pairs begin to dominate. The Fermi surface of
the normal-state has two sheets $k_{F0}\pm\frac{1}{\hbar v_{F}}\sqrt{h^{2}+\lambda^{2}+2h\lambda\hat{k}_{y}}$
where $k_{F0}$ refers to $h=0$ and $\lambda=0$. In both limits
$h\ll\lambda$ and $h\gg\lambda$, the Fermi surface can be approximated
by two circles of radii $k_{F0}\pm\frac{1}{\hbar v_{F}}\sqrt{h^{2}+\lambda^{2}}$
centered at $\left(0,\pm\frac{1}{\hbar v_{F}}\,\frac{h\lambda}{\sqrt{h^{2}+\lambda^{2}}}\right)$
(see Fig. \ref{fig:TwoVectors}). The optimal pairing is with respect
to these shifted centers, $\frac{{\bf q}}{2}=\left(0,\pm\frac{1}{\hbar v_{F}}\,\frac{h\lambda}{\sqrt{h^{2}+\lambda^{2}}}\right)$.
This result remains valid also for $h\sim\lambda$, as can be seen
from Fig. \ref{fig:EvolutionOfSecondSolution}.

Of special interest is the tricritical point (TP) at which the normal
phase merges with the superconducting phases with $Q=0$ and $Q\neq0$.
When $\lambda=0$ the TP is at $T_{{\rm tri}}(\lambda=0)/T_{c}=0.56$
and $h_{{\rm tri}}(\lambda=0)/\Delta_{0}=0.62$ \cite{SaintJamesBook,Maki64}.
When $\lambda$ increases from $\lambda=0$ the homogeneous superconducting
state is initially stabilized and $T_{{\rm tri}}/T_{c}$ decreases
until at $\lambda_{c2}\simeq0.722\Delta_{0}$ it reaches a minimum
of $T_{{\rm tri}}(\lambda_{c2})/T_{c}=0.531$ while $h_{{\rm tri}}(\lambda_{c2})/\Delta_{0}=0.675$.
For $\lambda>\lambda_{c2}$, both $T_{{\rm tri}}(\lambda)/T_{c}$
and $h_{{\rm tri}}(\lambda)$ increase continuously with $\lambda$
\cite{SupplTriCrit}.

In summary we have derived analytic expressions for the critical magnetic
field $h(T)$ of ultrathin films for all sizes of the Rashba spin-orbital
interaction energy in units of $\Delta_{0}$. With increasing ratio
$h/\lambda$ Cooper pairing changes from intraband to interband electron
states. We found a discontinuous jump of the pairing momentum ${\bf Q}$
taking place at a critical $\lambda_{c}$. For $\lambda>\lambda_{c}$
a dimensional cross-over of $h(T)$ takes place from two to one dimension.

\section*{Acknowledgement}

We would like to thank Dr. Yuri Ovchinnikov for helpful discussions.
SJ acknowledges funding by the School for Contacts in Nanosystems.
The work of GZ was performed in part at the Aspen Center for Physics,
which is supported by National Science Foundation grant PHY-1066293.


\end{document}


\widetext
\begin{center}
\textbf{\large Supplemental material for\\[2ex] ``Critical magnetic field of ultrathin superconducting films and Interfaces''}
\end{center}

\author{G. Zwicknagl$^1$, S. Jahns$^1$, and P. Fulde$^2$}
\affiliation{$^{1}$Institut f\"ur Mathematische Physik, Technische Universit\"at Braunschweig, 38106 Braunschweig, Germany\\
$^2$Max-Planck-Institut f\"ur Physik komplexer Systeme, 01187 Dresden, Germany}

\date{\today}

\maketitle
\begin{figure}[h]
\includegraphics[width=0.9\textwidth]{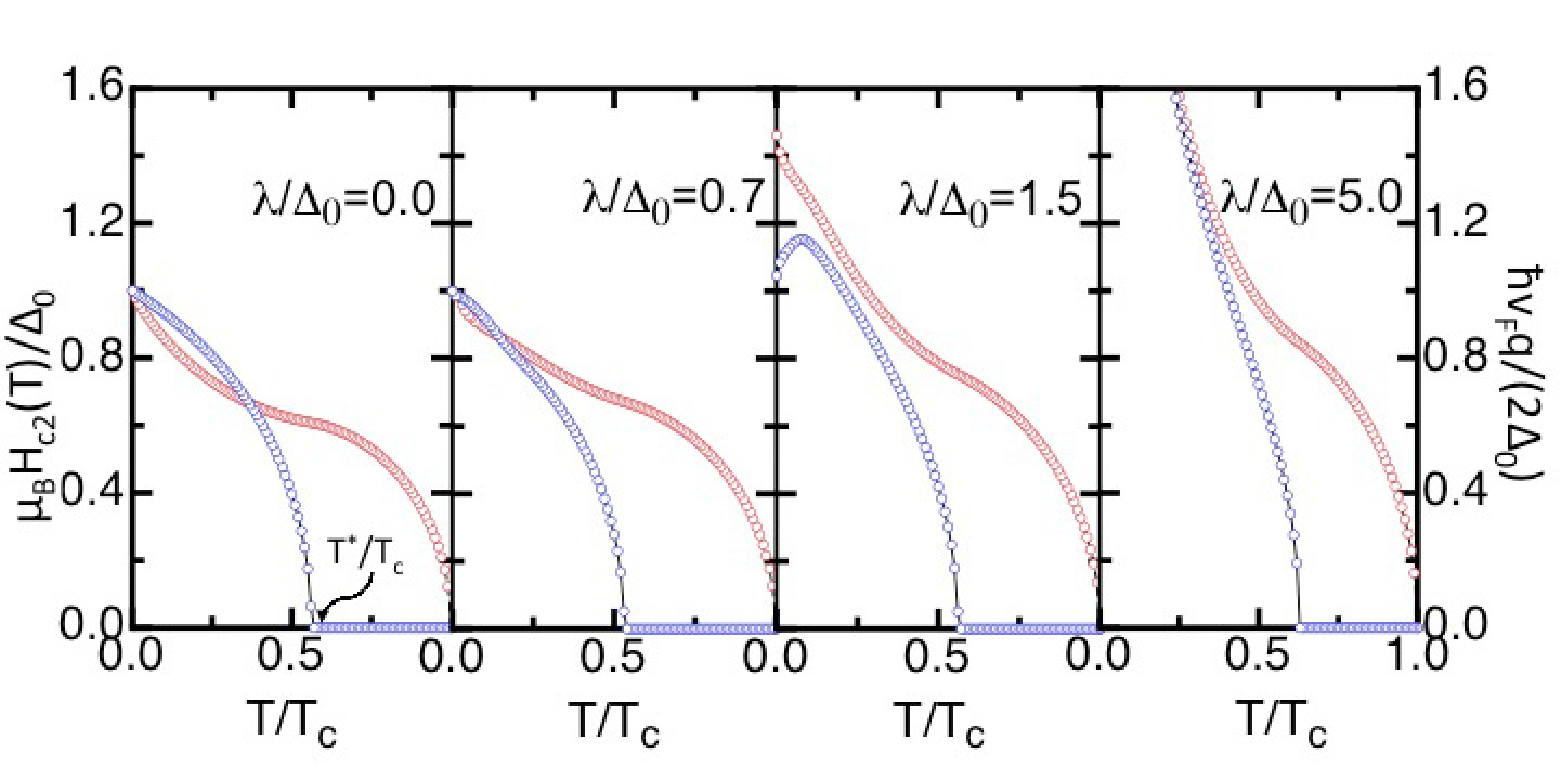}
\caption{{\footnotesize Critical field $h_{c2} (T)/\Delta_0$ (red) and pairing momentum $\hbar q v_F/2 \Delta_0$ (blue) obeying Eq. (6) as function of $T/T_c$. One notices that at $T = 0$ and $\lambda < \lambda_{\rm crit} =  \Delta_0/\sqrt{2}$ it is $h_{c2}(0) / \Delta_0 = 1$ and $\hbar v_F q / (2 \Delta_0) = 1$. For $\lambda > \lambda_{\rm crit}$ both quantities increase and cross over to one-dimensional behavior. One also notices that the tricritical temperature $T^*$, characterized by the onset of $q(\lambda) \neq 0$ decreases for $\lambda < \lambda_c$ and increases for $\lambda > \lambda_{c}$.}}
\label{fig:S01}
\end{figure}

\section{Variation of the upper critical field with temperature}
We present additional data obtained by evaluating Eq. (6) and illustrating the influence of the Rashba spin-orbit interaction on the upper critical field as shown in Fig.\ref{fig:S01}. The {\it qualitative} behavior of $h_{c2}\left(T\right)/\Delta_{0}=\mu_{B}H_{c2}\left(T\right)/\Delta_{0}$
is only weakly affected by the spin-orbit interaction. In particular,
we find in the vicinity of the transition temperature $T_c$
\begin{equation}
\frac{h_{c2}\left(T\right)}{2\pi k_{B}T_{c}}\sim\sqrt{1-\frac{T}{T_{c}}}~~~.
\end{equation}
The $\lambda$-dependent prefactor is derived by expanding Eq. (6)
to leading order in the the small quantities $\frac{h_{c2}\left(T\right)}{2\pi k_{B}T_{c}}$
and $\frac{T_{c}-T}{T_{c}}$
\begin{equation}
1-\frac{T}{T_{c}}=\left(\frac{h_{c2}\left(T\right)}{2\pi k_{B}T_{c}}\right)^{2}\frac{1}{2}\left\{ 7\zeta\left(3\right)+\frac{\mathrm{Re}\,\psi\left(\frac{1}{2}+i\frac{\lambda}{2\pi k_{B}T_{c}}\right)-\psi\left(\frac{1}{2}\right)}{\left(\frac{\lambda}{2\pi k_{B}T_{c}}\right)^{2}}\right\} 
\end{equation}
with the Riemann $\zeta$-function. This expression yields 
\begin{equation}
\frac{h_{c2}\left(T\right)}{2\pi k_{B}T_{c}}\simeq\sqrt{1-\frac{T}{T_{c}}\,}\begin{cases}
\sqrt{\frac{1}{7\zeta\left(3\right)}} & \quad;\quad\lambda\to0\\
\sqrt{\frac{2}{7\zeta\left(3\right)}} & \quad;\quad\lambda\to\infty
\end{cases}\quad.
\end{equation}

The sensitivity of the upper critical field for weak spin-orbit interaction
$\lambda$, i.e., $\left(\frac{\lambda}{\Delta_{0}}\,\frac{dh_{c2}}{d\lambda}\right)_{\lambda=0}$
, will be discussed below. The influence of a very strong spin-orbit interaction,
$\lambda\gg h,\,\Delta_{0}$, on the other hand, can be accounted
for by an effective anisotropic $g$-factor. 

For all values of the ratio $\lambda/\Delta_0$, in high magnetic fields a modulated superconducting phase with finite-momentum Cooper pairs is predicted to form at low temperatures $T<T^*\left(\lambda\right)$. The variation with $\lambda$ of the tricritical temperature $T^*\left(\lambda\right)$ is discussed below.

It is clearly seen that the discontinuous change in the $T=0$ pairing momentum at the critical value $\lambda_c=\Delta_0/\sqrt{2}$ leads to a non-monotonic variation of $q\left(T\right)$ with $T$. It reflects the competition between intra- and inter-band pairing in the ground state. 

One notices a steep increase of the pairing momentum and of $H_{c2} (T)$ when $\lambda/\Delta_0$ exceeds unity and an increase of $\hbar v_F q/(2 \Delta_0)$ at $T = 0$ for $\lambda > \lambda_c = \Delta_0/\sqrt{2}$. The dramatic increase in $H_{c2} (T)$ when $\lambda > \lambda_c$ should be experimentally observable.

\section{Inter- vs. intra-band pairing}

The short-ranged attractive interaction between the quasiparticles
leads to the formation of spin singlet Cooper pairs in the bulk reference
superconductor. In a thin film which is subject to both Rashba spin-orbit interaction and an in-plane magnetic field the pair wave functions
will contain both intra- and inter-band contributions
\begin{eqnarray}
\frac{1}{\sqrt{2}}\left(\left|{\bf k}\uparrow;-{\bf k}\downarrow\right\rangle -\left|{\bf k}\downarrow;-{\bf k}\uparrow\right\rangle \right) & = & u_{intra}\left(h,\lambda;\varphi\right)\,\frac{1}{\sqrt{2}}\left(\left|{\bf k}+;-{\bf k}+\right\rangle -\left|{\bf k}-;-{\bf k}-\right\rangle \right)
\nonumber \\
 &  & +u_{inter}\left(h,\lambda;\varphi\right)\,\frac{1}{\sqrt{2}}\left(\left|{\bf k}+;-{\bf k}-\right\rangle -\left|{\bf k}-;-{\bf k}+\right\rangle \right)
\end{eqnarray}
where the end symbols $+$, $-$ label the two Fermi surfaces (see Fig. 5). The anisotropic weights
\begin{eqnarray}
\left|u_{intra}\left(h,\lambda;\varphi\right)\right|^{2} & = & \frac{1}{2}\left\{ 1+\frac{h^{2}-\lambda^{2}}{\sqrt{\left(h^{2}+\lambda^{2}\right)^{2}-4h^{2}\lambda^{2}\sin^{2}\varphi}}\right\} \nonumber \\
\left|u_{inter}\left(h,\lambda;\varphi\right)\right|^{2} & = & \frac{1}{2}\left\{ 1-\frac{h^{2}-\lambda^{2}}{\sqrt{\left(h^{2}+\lambda^{2}\right)^{2}-4h^{2}\lambda^{2}\sin^{2}\varphi}}\right\} 
\end{eqnarray}
depend on the ratio $\bar{h}=h/\lambda$. The angular averages for
isotropic s-wave pairing as well as for $d_{x^{2}-y^{2}}$-pairing
with nodal and anti-nodal direction parallel to the applied magnetic
field are easily evaluated in closed form 
\begin{eqnarray*}
\left\langle \left|u_{intra/inter}\left(h,\lambda;\varphi\right)\right|^{2}\right\rangle _{s-wave} & = & \frac{2}{\pi}\int_{0}^{\pi/2}\left|u_{intra/inter}\left(h,\lambda;\varphi\right)\right|^{2}\nonumber \\
 & = & \frac{1}{2}\left\{ 1\pm\frac{\bar{h}^{2}-1}{\bar{h}^{2}+1}\:\frac{2}{\pi}K\left(\frac{2\bar{h}}{\bar{h}^{2}+1}\right)\right\} \\[3ex]
\left\langle \left|u_{intra/inter}\left(h,\lambda;\varphi\right)\right|^{2}\right\rangle _{d-wave;anti} & = & \frac{2}{\pi}\int_{0}^{\pi/2}\left|u_{intra/inter}\left(h,\lambda;\varphi\right)\right|^{2}2\cos^{2}2\varphi\nonumber \\
 & = & \frac{1}{2}\left\{ 1\pm\frac{\bar{h}^{2}-1}{\bar{h}^{2}+1}\,\frac{2}{\pi}\frac{1}{3\bar{h}^{4}}\right.\times\nonumber \\
 &  & \left.\left[\left(1+4\bar{h}^{4}+\bar{h}^{8}\right)K\left(\frac{2\bar{h}}{\bar{h}^{2}+1}\right)-\left(1+\bar{h}^{2}\right)^{2}\left(1+\bar{h}^{4}\right)E\left(\frac{2\bar{h}}{\bar{h}^{2}+1}\right)\right]\right\} \\[3ex]
\left\langle \left|u_{intra/inter}\left(h,\lambda;\varphi\right)\right|^{2}\right\rangle _{d-wave;nodal} & = & \frac{2}{\pi}\int_{0}^{\pi/2}\left|u_{intra/inter}\left(h,\lambda;\varphi\right)\right|^{2}2\sin^{2}2\varphi\nonumber \\
 & = & \frac{1}{2}\left\{ 1\pm\frac{\bar{h}^{2}-1}{\bar{h}^{2}+1}\,\frac{2}{\pi}\frac{\left(1+\bar{h}^{2}\right)^{2}}{3\bar{h}^{4}}\right.\times\nonumber \\
 &  & \left.\left[\left(1+\bar{h}^{4}\right)E\left(\frac{2\bar{h}}{\bar{h}^{2}+1}\right)-\left(1-\bar{h}^{2}\right)^{2}K\left(\frac{2\bar{h}}{\bar{h}^{2}+1}\right)\right]\right\} \nonumber \\
 &&
\end{eqnarray*}
where $K$ and $E$ denote the complete elliptic integrals \cite{Gradshteyn}
\begin{eqnarray}
K\left( k \right)& = & \int_0^{\pi/2}\,\frac{d\varphi}{\sqrt{1-k^2\sin^2 \varphi}} \nonumber \\
 & & \nonumber \\
E\left( k \right) & = & \int_0^{\pi/2}\,d\varphi\,\sqrt{1-k^2\sin^2 \varphi}  \quad .
\end{eqnarray}

\section{Sensitivity of $H_{c2}\left( T \right)$ with respect to small Rashba Spin-Orbit interaction}
\begin{figure}[h]
\includegraphics[width=0.6\textwidth]{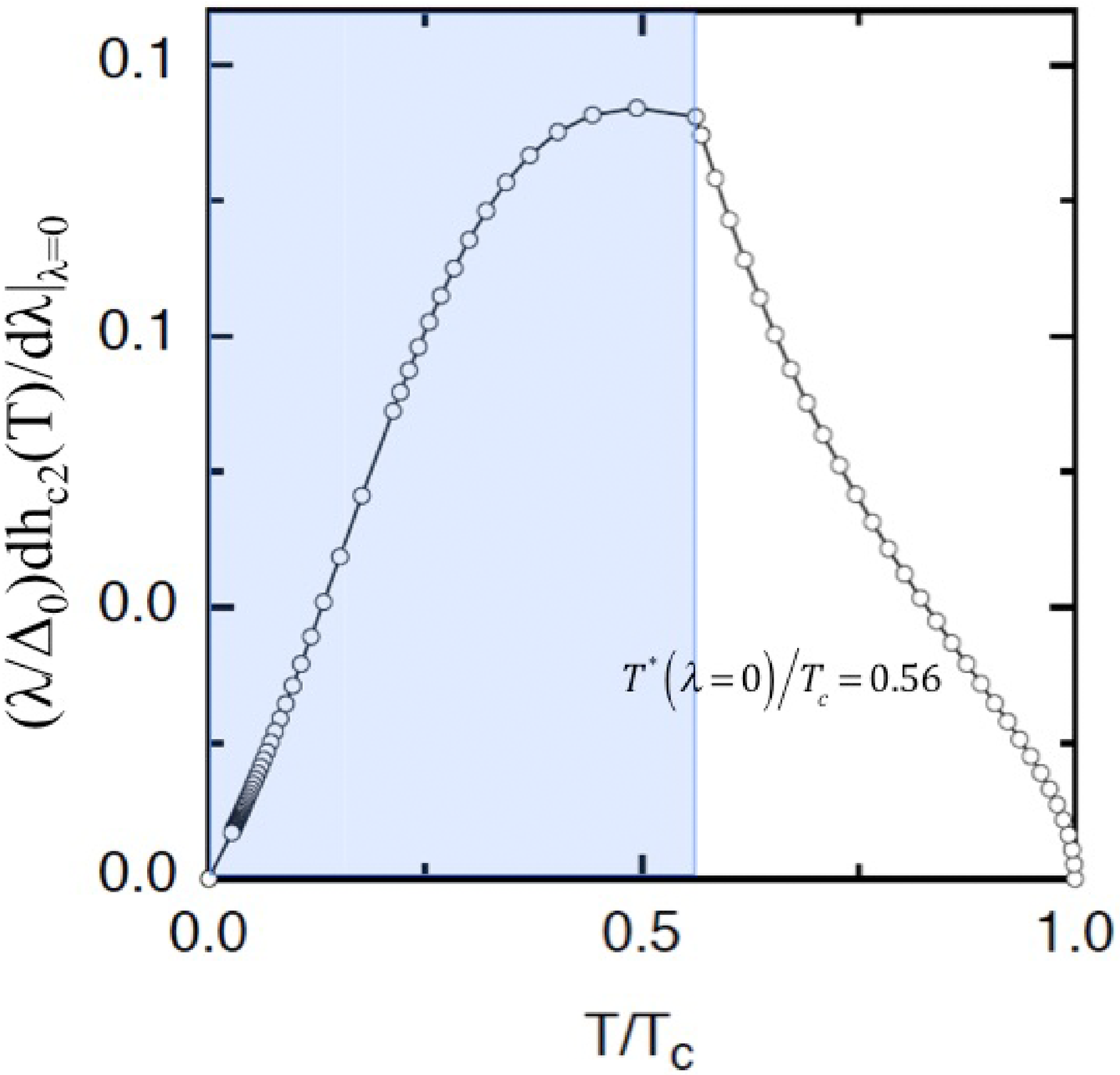}
\caption{{\footnotesize Sensitivity of $h_{c2} (T)$ with changing spin-orbit interaction as function of temperature. A sudden change in the behavior at $T = T^* (\lambda = 0)$ is noticed.}}
\label{fig:S03}
\end{figure}
It is also instructive to study the sensitivity of the critical field $h_{c2} (T)$ with respect to changes in $\lambda$. It is found that there is a significant change at the tricritical temperature $T^* (\lambda=0)/T_c = 0.531$ (see Fig. \ref{fig:S03}). Starting point is here Eq. (4). 
We expand the kernel $K$ for small values of $\lambda$
\begin{equation}
\left( \frac{dK}{d \lambda^2} \right)_{\lambda = 0} = \left( \frac{\partial K}{\partial \lambda^2} + \frac{\partial K}{\partial H} \frac{\partial H}{\partial \lambda^2} + \nabla_{\bf q} K \cdot \frac{\partial {\bf q}}{\partial \lambda^2} \right)_{\lambda = 0}~~.
\label{eq:S1}
\end{equation}
For the upper critical field $H_{c2}$ we have 
\begin{equation}
\nabla_{\bf q} K=0
\end{equation}
which implies
\begin{equation}
\left( \frac{\partial K}{\partial \lambda^2} \right)_{H_{c2}, {\bf q}, \lambda = 0} + \left( \frac{\partial K}{\partial H}\right)_{H_{c2}, {\bf q}, \lambda = 0} \left( \frac{\partial H_{c2}}{\partial \lambda^2} \right)_{\lambda = 0} = 0~~.
\label{eq:S2}
\end{equation}
From this we obtain
\begin{equation}
\left( \frac{\partial H_{c2}}{\partial \lambda^2} \right)_{\lambda = 0} = - \frac{\left( \frac{\partial K}{\partial \lambda} \right)_{H_{c2}, {\bf q}, \lambda = 0}}{\left( \frac{\partial K}{\partial H} \right)_{H_{c2}, {\bf q}, \lambda = 0}} ~~.
\label{eq:S3}
\end{equation}
From Eq. (6) we obtain
\begin{equation}
\left( \frac{\partial K}{\partial h} \right)_{H_{c2}, {\bf q}, \lambda = 0} = - \frac{1}{2 \pi k_B T} \left< \mathrm{Im}\, \psi' \left( \frac{1}{2} + i \frac{h_{c2} + \frac{\hbar}{2} {\bf v}_F \cdot {\bf q}}{2 \pi k_B T} \right) \left| \mathcal{Y} \left( \hat{\bf k} \right) \right|^2  \right>_{FS} 
\label{eq:S4}
\end{equation}
and
\begin{eqnarray}
\left( \frac{\partial K}{\partial \lambda^2} \right)_{H_{c2}, {\bf q}, \lambda = 0} & = & - \frac{1}{h^2_{c2}} \left< \left[  \mathrm{Re}\, \psi \left( \frac{1}{2} + i \frac{h_{c2} + \frac{\hbar}{2} {\bf v}_F \cdot {\bf q}}{2 \pi k_B T} \right) -  \mathrm{Re}\, \psi \left( \frac{1}{2} + i \frac{\frac{\hbar}{2} {\bf v}_F \cdot {\bf q}}{2 \pi k_B T} \right) \right] \left| \mathcal{Y} \left( \hat{\bf k} \right) \right|^2 \right>_{FS} \nonumber\\
&& - \frac{1}{h_{c2}} \frac{1}{2\pi k_B T} \left<  \mathrm{Im}\, \psi' \left( \frac{1}{2} + i \frac{h_{c2} + \frac{\hbar}{2} {\bf v}_F \cdot {\bf q}}{2 \pi k_B T} \right) \cos^2 \phi \left|  \mathcal{Y}\left( \hat{\bf k} \right) \right|^2 \right>_{FS}~~.
\label{eq:S5}
\end{eqnarray}
The notation is the same as in Figure 1. Numerical evaluations result are shown in Fig.\ref{fig:S03}. 

The findings can be summarized as follows: First, the change $\left. \frac{\lambda}{\Delta_0}\frac{dh_c2}{d\lambda}\right\vert$ is positive over the entire temperature range. This is to be expected since the spin-orbit interaction reduces the magnetic polarization which, in turn, limits spin singlet pairing. Second, the critical temperature $T_c$ of an isotropic s-wave superconductor is not affected by weak spin-orbit interaction $\lambda \ll E_F$, in agreement with Anderson's theorem. Third, the behavior changes at the tricritical point $T^*$ where the normal state merges with the homogeneous and a modulated superconducting state. In this regime, the upper critical field is highly sensitive to spin-orbit interaction. Finally, the upper critical field  of the quasi-2D superconductor remains unaffected at $T=0$ by weak spin-orbit interaction.

\section{Tricritical point}
\begin{figure}[h]
\includegraphics[width=0.45\textwidth]{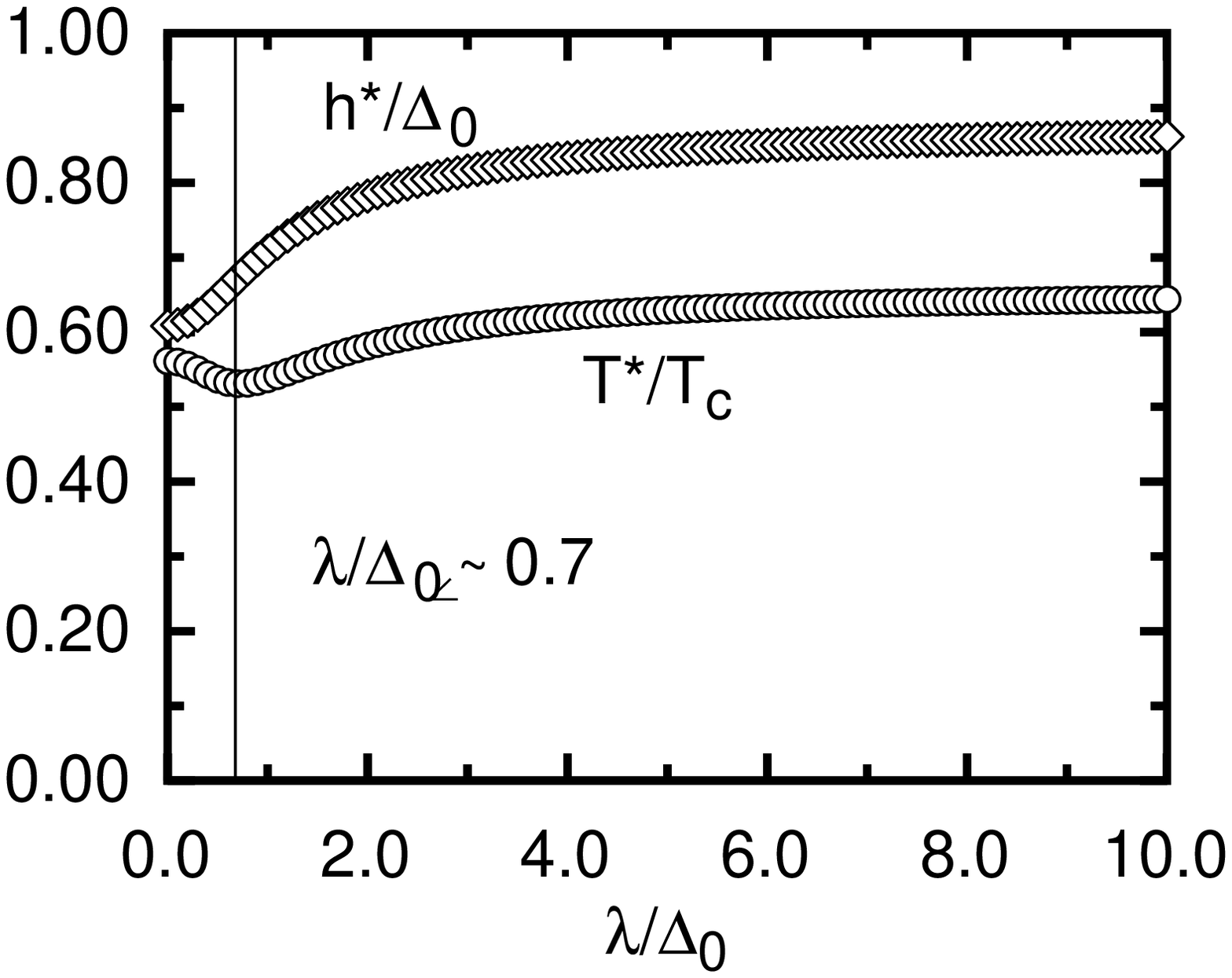}
\caption{{\footnotesize Tricritical temperature $T^*$ and magnetic field $h^*$ in units of $\Delta_0$ as function of $\lambda/\Delta_0$. One notices an initial decrease of $T^*$ from its value at $\lambda = 0$. The minimum of $T^*/\Delta_0$ is reached at $\lambda \sim 0.7$.}}
\label{fig:S02}
\end{figure}

A special finding has been the initial decrease of $T^*/T_c$ with $\lambda \neq 0$ from its value $T^* (\lambda = 0)/T_c = 0.56$. The minimum value is reached for $\lambda = \lambda_{c2} \simeq 0.7 \Delta_0$. Beyond this point, $T^*/T_c$ as well as the critical magnetic field $h^*$ at $T^*$ increase continuously with $\lambda$. This is demonstrated in Fig. \ref{fig:S02}.

\section{Comparison with spin-orbit scattering from impurities}
It is interesting to compare Eq. (6) for ${\bf q} = 0$ with the corresponding expression when the spin-orbit interaction is due to scattering of electrons by impurities or surfaces instead of the Rashba interaction. These processes do not conserved the momentum of the electrons. They conserve time-reversal symmetry though. Therefore the single electron states are Kramers degenerate and Cooper pairing takes place between these. The spin is no longer a good quantum number and therefore the spin susceptibility of the superconducting ground state is finite and depends on the spin-orbit scattering rate $\tau^{-1}_{\rm SO}$. If
\begin{equation}
v \left( {\bf k}, {\bf k}' \right) = \frac{iv_{\rm SO}}{k^2_F} \left[ {\bf k} \times {\bf k}' \right]_\sigma
\label{eq:S6}
\end{equation}
denotes the spin-orbital part of the scattering potential of an impurity the scattering rate is
\begin{equation}
\frac{1}{\tau_{\rm SO}} = n_i \frac{N(0)}{2} \int d \Omega \left| v_{\rm SO} \right|^2 \sin^2 \vartheta
\label{eq:S7}
\end{equation}
where $\vartheta$ is the angle between ${\bf k}$ and ${\bf k}'$ and $n_i$ is the impurity concentration. As usual $N(0)$ is the electron density of states in the normal state per spin direction. In addition to the spin-orbit scattering rate there is also an isotropic Coulomb scattering rate $\tau^{-1}$ when impurities are present. This rate is usually much larger than $\tau^{-1}_{\rm SO}$ so that often the dirty limit $\tau \Delta_0 \ll 1$ is assumed. 

The mathematical technique of treating a spin-orbit component in the impurity scattering amplitude is due to Abrikosov and Gorkov \cite{Abrikosov62}.
When we introduce the dimensionless quantities $b = (3 \tau_{\rm SO} \Delta_0)^{-1}$ and $\tilde{h} = h/\Delta_0$ (with $h = \mu_B H_{c2}$) we obtain by following \cite{Maki66}, or alternatively \cite{Werthamer66}:
\begin{eqnarray}
\ln T/{T_c} & = & \psi (1/2) - \frac{1}{2} \left( 1 + \frac{b}{\left( b^2 -\tilde{h}^2 \right)^{1/2}} \right) \psi \left( \frac{1}{2} + \frac{\Delta_0}{2 \pi T} \left( b - \left( b^2 - \tilde{h}^2 \right)^{1/2} \right) \right)\nonumber\\
&& - \frac{1}{2} \left( 1 - \frac{b}{\left( b^2 - \tilde{h}^2 \right)^{1/2}} \right) \psi \left( \frac{1}{2} + \frac{\Delta_0}{2 \pi T} \left( b + \left( b^2 - \tilde{h}^2 \right)^{1/2} \right)^{1/2} \right)~~.
\label{eq:S8}
\end{eqnarray}
The similarity but also the difference to Eq. (6) with ${\bf q} = 0$ is easily recognized. The difference between the two equations is due to the different stage at which the averaging over the direction of the electron momentum is done when Cooper pairs form. In (6) this average is done after the electron-phonon interaction is treated while in (\ref{eq:S8}) due to $\tau \Delta_0 \ll 1$ it is done before the electrons interact with the phonons and pair up \cite{Maki66,Werthamer66}. The dependence of $H_{c2} (T)$ as obtained from (\ref{eq:S8}) differs strongly from the one of (6). The equation simplifies further when $b \gg 1$. In this case (\ref{eq:S8}) reduces to
\begin{equation}
\ln \frac{T}{T_c} = \psi \left( \frac{1}{2} \right) - \psi \left( \frac{1}{2} + \frac{\alpha}{2 \pi T}  \right)
\label{eq:S9}
\end{equation}
with $\alpha = \tilde{h}^2/(2b)$. This type of equation is well known from the theory of Abrikosov and Gorkov \cite{Abrikosov60} for magnetic impurities in superconductor. It applies also to numerous other situations when time-reversal symmetry breaking perturbations are acting on spin singlet Cooper pairs like here the Zeeman field. A number of different features following from (\ref{eq:S8}) have been discussed in detail in \cite{Fulde73} and need not be repeated here.